\documentclass[onecolumn,12pt,draftclsnofoot]{IEEEtran}
\usepackage{graphicx}
\usepackage{array}
\usepackage{epstopdf}
\usepackage{mdwmath}
\usepackage{mdwtab}
\usepackage{eqparbox}
\usepackage{algorithmic}
\usepackage[cmex10]{amsmath}
\usepackage{setspace}
\usepackage{algorithm,caption}
\usepackage{empheq}
\usepackage{subcaption}
\usepackage{cases}
\usepackage{soul}
\usepackage{color}
\usepackage{cite}
\usepackage{amssymb}
\usepackage{multirow}
\usepackage{url}
\allowdisplaybreaks
\ifCLASSINFOpdf

\else

\fi
\begin{document}
%
\title{An Overview of Sustainable Green 5G Networks }
%
%
%

\author{\IEEEauthorblockN{Qingqing Wu, Geoffrey Ye Li,   Wen Chen, Derrick Wing Kwan Ng, and Robert Schober
\thanks{Qingqing Wu and Geoffrey Ye Li are with the School of Electrical and Computer Engineering, Georgia Institute of Technology, USA,
email:\{qingqing.wu, liye\}@ece.gatech.edu. Wen Chen is with Department of Electronic Engineering, Shanghai Jiao Tong University, China,
email: wenchen@sjtu.edu.cn. Derrick Wing Kwan Ng is with the School of Electrical Engineering and Telecommunications, The University of New South Wales, Australia, email: w.k.ng@unsw.edu.au.  Robert Schober is with the Institute for Digital Communications, Friedrich-Alexander-University Erlangen-N$\rm{\ddot{u}}$rnberg, email: robert.schober@fau.de.}}  }
\maketitle


\begin{abstract}
 The stringent requirements of a 1,000 times increase in data traffic and one millisecond round trip latency have made limiting the potentially tremendous ensuing energy consumption one of the most challenging problems for the design of the upcoming fifth-generation (5G) networks.
To enable sustainable 5G networks,
new technologies have been proposed to improve the system energy efficiency and alternative energy sources are introduced
to reduce our dependence on traditional fossil fuels. In particular, various 5G techniques target the reduction of the energy consumption without sacrificing the quality-of-service.
Meanwhile, energy harvesting technologies, which enable communication transceivers to harvest energy from various renewable resources and ambient radio frequency signals for communication, have drawn significant interest from both academia and industry.
In this article, we provide an overview of the latest research on both green 5G techniques and energy harvesting for communication. In addition,  some technical challenges and potential research topics for realizing sustainable green 5G networks are also identified.
\end{abstract}

%
%


\IEEEpeerreviewmaketitle

\section{Introduction}
The fifth-generation (5G) wireless networks will support up to a 1,000-fold increase in capacity compared to the existing networks.  It is anticipated to connect at least 100 billion devices worldwide with approximately 7.6 billion mobile subscribers due to the tremendous popularity of smartphones, electronic tablets, sensors, etc. and provide an up to 10 Gb/s individual user experience \cite{andrews2014will}.  Along with the dramatic traffic explosion and device proliferation, 5G wireless networks also have to integrate human-to-machine and machine-to-machine communications in order to facilitate more flexible networked social information sharing aiming for one million connections per square kilometer. Consequently, sensors, accessories, and  tools are expected to become wireless  communication entities exchanging information, giving rise to the well-known ``Internet of Things (IoT)''. 
 With such tremendously expanding demand for wireless communications in the future, researchers are currently looking for viable solutions to meet the stringent throughput requirement. In this regard, three paradigms have emerged:
\begin{itemize}
  \item Reduce the transmitter-receiver (Tx-Rx) distance and improve frequency reuse: ultra-dense networks (UDNs) and device-to-device (D2D) communications;
  \item Exploit unused and unlicensed spectrum: millimeter wave (mmWave) communications and Long Term Evolution (LTE) in unlicensed spectrum (LTE-U);
  \item Enhance spectral efficiency (SE) by deploying a massive amount of antennas: massive multiple-input multiple-output (M-MIMO).
\end{itemize}

The technologies listed above increase the system throughput from three different angles.
However, the performance gains introduced by these technologies do not come for free. 
For example, M-MIMO exploits a large number of antennas for potential multiplexing and diversity gains at the expense of an escalating circuit power consumption in the radio frequency (RF) chains which scales linearly with the number of antennas. In addition, power-hungry transceiver chains and complex signal processing techniques are needed for reliable mmWave communications to operate at extremely high frequencies.
In light of this, the network energy consumption may no longer be sustainable and designing energy-efficient 5G networks is critical and necessary.
In fact, in recent years, energy consumption has become a primary concern in the design and operation of wireless communication systems motivated by the desire to lower the operating cost of the base stations (BSs), prolong the lifetime of the user terminals (UTs), and also protect the environment.
 As a result, energy efficiency (EE), measured in bits-per-Joule, has emerged as a new prominent figure of merit and has become the most widely adopted green design metric for wireless communication systems \cite{chen2011fundamental}.

\begin{figure*}[!t]
\setlength{\abovecaptionskip}{0pt}
\setlength{\belowcaptionskip}{0pt}
\centering
\includegraphics[width=0.8\textwidth]{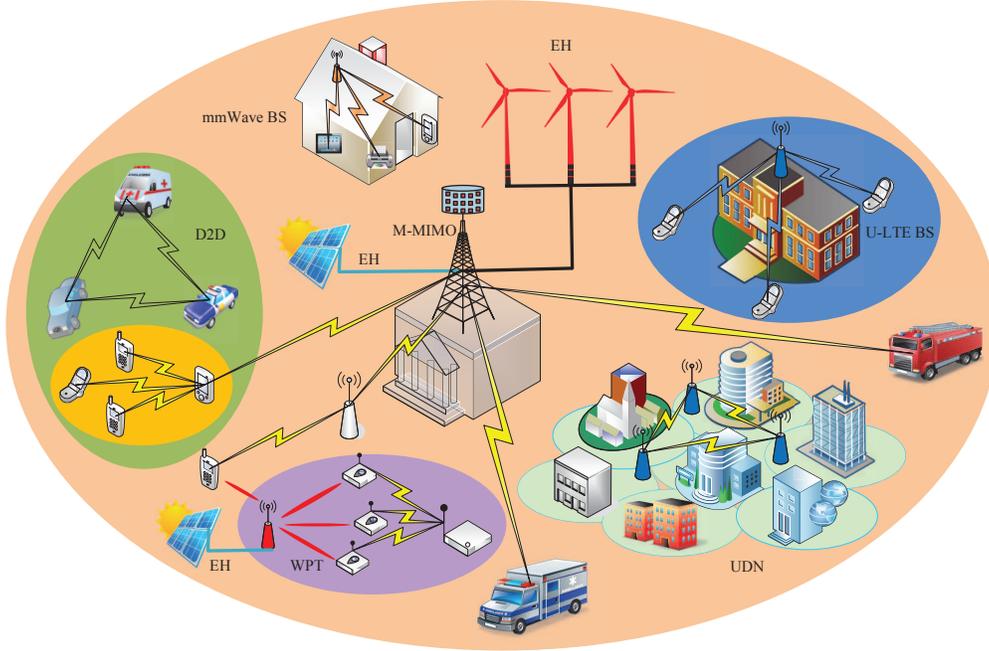}
\centering
\caption{An ecosystem of disruptive technologies for 5G wireless networks. }\label{system}
\centering
\end{figure*}

Meanwhile, it should also be noted that EE cannot be improved infinitely by only applying spectral efficient communication technologies,  due to the constraint imposed by Shannon capacity bound as well as the non-negligible circuit power consumption. Consequently,  even with the aforementioned 5G paradigms for improving
the power consumption will still grow because of the explosive future data rate requirements. Therefore, improving EE can only alleviate the power consumption problem to a certain extent and is insufficient for enabling sustainable 5G communications.
Hence, energy harvesting technologies, which allow BSs and devices to harvest energy from renewable resources (solar, wind, etc.) and even RF signals (television signals, interference signals, etc.), have received significant attention recently.
 They provide green energy supply solutions for powering different components of wireless communication networks.
 Therefore, integrating energy harvesting technologies into future wireless networks is imperative \cite{gunduz2014designing}.
Gaining a better understanding of the specific challenges of energy harvesting technologies and alternative energy sources will pave the way for incorporating them smoothly into the upcoming 5G wireless network planning, design, and deployment.
This article focuses on the state-of-the-art of energy harvesting and green 5G technologies, which will create an ecosystem of  5G wireless networks as shown in Fig. \ref{system}.

The rest of this article is organized as follows. After briefly introducing the tradeoff between EE and spectral efficiency (SE) in Section II, we discuss green 5G technologies from the perspectives of enlarging spectrum availability, shortening Tx-Rx distances, and enhancing the spatial degrees of freedom in Sections III, IV, and V, respectively. Then, energy harvesting technologies for 5G systems are discussed in Section IV, and conclusions are drawn in Section VI.
\begin{table*}[!t]
\caption{Comparison of green 5G technologies.}\label{table1}
\centering
\newcommand{\tabincell}[2]{\begin{tabular}{@{}#1@{}}#2\end{tabular}}
\small
\begin{tabular}{|p{2cm}|p{2cm}|p{1.5cm}|p{2.23cm}|p{2.15cm}|p{2.15cm}|}
  \hline
Technology &High EE at& Coverage & Transmit Power & Circuit Power & Signalling Overhead \\ \hline
   \tabincell{l}{mmWave} & BS and UT & 200 m & Low & High  & High\\ \hline
LTE-U&  BS and UT & 500 m & Moderate & Moderate  &  Moderate \\ \hline
   \tabincell{l}{UDNs}&UT & 10-200 m& Low &High & High\\ \hline
  \tabincell{l}{D2D}& BS& 2-100 m & Low & Low  & Moderate \\ \hline
   M-MIMO& \tabincell{l}{UT } & 1000 m &Low & High  & High\\ \hline
\end{tabular}
\end{table*}

\begin{figure}[!t]
\setlength{\abovecaptionskip}{0pt}
\setlength{\belowcaptionskip}{0pt}
\centering
\includegraphics[width=0.45\textwidth]{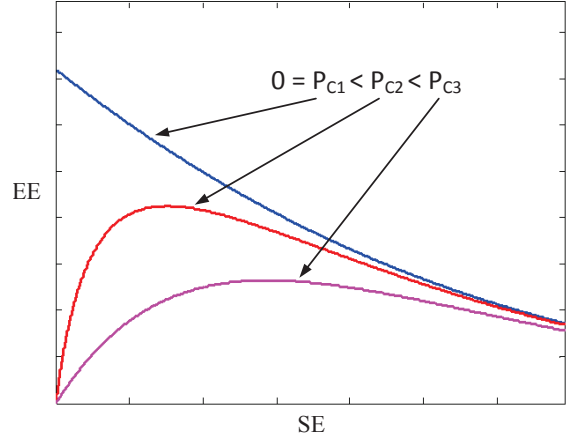}
\caption{Fundamental tradeoff between EE and SE.}\label{tradeoff}
\end{figure}


\section{General EE-SE Tradeoff}
In general, the system achievable throughput or SE can be expressed as
\begin{align}\label{SE}
{\rm{SE}} =K\times B\times N\times \log_2\bigg(1+{\rm{SINR}}(d)\bigg),
\end{align}
where $K$, $B$, $N$, and $d$ are the frequency reuse factor, the signal bandwidth, the number of spatial beams (spatial multiplexing factor), and a single link distance, respectively. $\rm{SINR}$ is the signal-to-interference-plus-noise ratio at the receiver that increases with decreasing $d$.
Correspondingly, the system EE can be expressed as
\begin{align}\label{EE}
{\rm{EE}} = \frac{K\times B\times N\times  \log_2\bigg(1+{\rm{SINR}}(d)\bigg)}{P_{\rm{t}} + P_{\rm{c}}},
\end{align}
where $P_{\rm{t}}$ and $P_{\rm{c}}$ are the consumed transmit and circuit powers, respectively.
From (\ref{SE}), the SE of wireless networks can be improved by increasing the frequency reuse factor (reducing the Tx-Rx distance), the signal bandwidth, and/or the number of spatial beams, which will be discussed in the subsequent three sections, respectively.
Table \ref{table1} compares various performance aspects of several 5G technologies. From the table, most of the 5G technologies enable the reduction of the transmit power at the expense of incurring additional circuit power consumption due to the required hardware expansion, sophisticated signal processing, etc.

For point-to-point link level communications, where $K$, $B$, $N$, and $d$ in (\ref{SE}) and (\ref{EE}) are fixed,  the relation between EE and SE relation can be analyzed using the approach in \cite{chen2011fundamental} which is illustrated in  Fig. \ref{tradeoff}. From the figure,
\begin{itemize}
  \item if the circuit power consumption is ignored, i.e., $P_{\rm{c}}=0$, the EE decreases monotonically with SE ;
  \item if the circuit power consumption is considered,  i.e., $P_{\rm{c}}>0$, the EE increases with increasing SE below a  threshold and decreases with increasing SE beyond the threshold;
    \item as the SE increases, regardless of the circuit power, the EE eventually converges to the same values as for $P_{\rm{c}}=0$, due to the domination of the transmit power;
        \item reducing the circuit power will enlarge the EE-SE tradeoff region.
\end{itemize}

Although observed from a single communication link, the fundamental tradeoff between EE and SE in Fig. \ref{tradeoff} carries over to more complicated systems employing the aforementioned 5G technologies. Besides, the EE and SE tradeoff can be achieved via different system configurations such as spectrum management, frequency reuse, spatial multiplexing, power allocation, etc. In addition, energy harvesting technologies can provide green energy, which allows 5G networks to possibly operate at higher SEs compared to the conventional energy limited networks. 


\section{Green 5G Technologies: Enlarging Spectrum availability}
Given the extreme shortage of available spectrum at traditional cellular frequencies, both mmWave (30 to 300 GHz) and LTE-U (5 GHz) communications aim at boosting the throughput by directly expanding the available radio spectrum for cellular networks, i.e., increasing $B$ in (\ref{SE}) and (\ref{EE}).


\subsection{ Millimeter Wave}
MmWave communications exploit the enormous chunks of spectrum available at mmWave frequencies and are expected to push the mobile data rates to gigabits-per-second for 5G wireless networks. However, the propagation conditions at the extremely high frequencies of the mmWave bands lead to several fundamental issues including high path attenuation, severe blockage effects, and atmospheric and rain absorption, which poses new challenges for providing guaranteed quality-of-service (QoS).
Thanks to the small wavelengths of mmWave bands, a large number of antenna elements can be integrated in a compact form at both the BSs and the UTs and leveraged to synthesize highly directional beams leading to large array gains. 
Energy-efficient mmWave communications comprises the following aspects.

\textbf{1) Energy-aware hybrid transceiver architectures:}
For mmWave frequencies, the conventional fully digital architecture employed for micro-wave frequencies, where each antenna is connected to a dedicated RF chain,
is unsustainable.
 In particular,  an excessive power consumption arises from the processing of massive parallel gigasamples per second data streams, leading to an energy  inefficient and expensive system. Thus, a direct approach to reduce power consumption in practice is to adopt analog beamforming. Specifically, one RF chain is connected to multiple antennas and the signal phase of each antenna is controlled by a network of analog phase shifters (PSs). However, pure analog beamforming with a single beamformer only supports single user and single data stream transmission, which does not fully exploit the potential multiplexing gains provided by multiple antennas. As such, hybrid architectures have been proposed as energy-efficient alternatives for mmWave communications \cite{han2015large}. Specifically, analog beamforming applies complex coefficients to manipulate the RF signals by means of controlling phase shifters and/or variable gain amplifiers and aims to compensate for the large path loss at mmWave bands, while digital beamforming is done in the form of digital precoding that multiplies a particular coefficient to the modulated baseband signal per RF chain to optimize capacity using various MIMO techniques.
Basically, two hybrid structures have been developed:
\begin{itemize}
  \item Fully-connected architecture: each RF chain is connected to all antennas.
  \item Partially-connected architecture: each RF chain is connected only to a subset of antennas.
\end{itemize}

From the energy-aware hardware design perspective, the fully-connected architecture requires thousands of PSs and introduces additional energy consumption, such as the high power compensation required for the giant phased radar and the insertion loss of the PSs. In contrast, for the partially-connected architecture, the number of  PSs needed  decreases by a factor equal to the number of RF chains and all signal processing is conducted at the subarray level. Hence, the system complexity as well as the circuit power consumption are significantly reduced, although at the expense of less flexibility in utilizing the spatial degrees of freedom. Nevertheless, for these structures, the hybrid precoding and combining schemes, the number of RF chains and antennas, and the transmit power can be separately or jointly optimized for customizing the system performance depending on the needs and requirements.
%

\textbf{2) Low resolution design:}
Besides hybrid structures, employing low resolution analog-to-digital converters (ADCs) at the receivers is an alternative approach towards energy-efficient designs.
The theoretical motivation is that the power dissipation of an ADC scales linearly with the sampling rate and exponentially with the number of bits per sample.  Furthermore, the data interface circuits connecting the digital components to the ADCs are usually power hungry and highly depend on the adopted resolution. Thus, the power consumption of high speed and high resolution ADCs becomes a critical concern and system performance bottleneck in mmWave systems employing a large number of antennas and very wide bandwidth.
This motivates the configuration of low resolution ADCs in practical systems,  especially for battery-constrained handheld devices.
 Yet, the characterization of the channel capacity taking into account low-resolution ADCs is in its infancy for mmWave communications and requires extensive research efforts. In addition, deploying a mix of high-resolution and low-resolution ADCs, i.e., mixed ADCs, is also a promising direction for power savings.

\subsection{LTE-U}
 LTE-U further increases the LTE capacity by enabling the LTE air interface to operate in unlicensed spectrum.
The unlicensed spectrum available in the 5 GHz band/WiFi band ($\geq 400$ MHz bandwidth) provides LTE-U with additional spectrum resources and has been launched in 3GPP Rel-13.

\textbf{Harmonious Coexistence:} The fundamental challenge of LTE-U is the coexistence of the LTE system and the incumbent unlicensed systems, such as WiFi systems. LTE employs a scheduling-based channel access mechanism, where multiple users can be served simultaneously by assigning them different resource blocks. In contrast, WiFi adopts contention-based channel access with a random backoff mechanism, where users are allowed to only access channels that are sensed as idle. Multi-user transmission with centralized scheduling enables LTE to make better use of the unlicensed band and also to achieve a higher EE than WiFi since the channel sensing and backoff time in WiFi lead to a waste of spectrum resources.
 However, if left unconstrained, LTE-U transmission may generate continuous interference to WiFi systems such that the channel is detected busy most of the time \cite{zhang2015lte}. This will lead to unceasing backoff times for the WiFi links and incur high energy consumption and low EE for the WiFi users even though no WiFi data is transmitted. Therefore, intelligent modifications to the resource management in the unlicensed band becomes critical to the harmonious coexistence. So far, two coexistence methods have been proposed in contributions to 3GPP:
 \begin{itemize}
\item Listen before talk (LBT): LTE-U devices are required to verify whether the channel is occupied by WiFi systems before transmitting.
\item Duty cycling: The LTE signal is periodically turned on and off, occupying and vacating the channel for periods of time without verifying the channel occupancy before transmitting.
 \end{itemize}
 Clearly, contention-based WiFi systems inherently follow the LBT protocol, which is deemed to facilitate a fair coexistence. However, an increasing number of contending devices will lead to high transmission collision probabilities and thereby limit the system performance. In light of this, the duty cycling protocol seems to enable a more efficient utilization of the unlicensed band.  However, duty cycling assumes that the carriers are in charge of on-and-off scheduling, which contradicts the common property nature of the unlicensed spectrum. This is a drawback as the carriers are under no real obligation to provide a decent time window for Wi-Fi, . Although there is no globally accepted coexistence protocol yet, one popular consensus  is that ``LTE-U is often a better neighbor to Wi-Fi than Wi-Fi itself".

\section{Green 5G Technologies: Shortening Tx/Rx distance}
Both UDNs and D2D communications boost the SE via shortening the distances between Txs and Rxs.  Short range communication has the dual benefits of providing high-quality links and a high spatial reuse factor compared to the current wireless networks, i.e.,  increasing $K$ and decreasing $d$ in (\ref{SE}) and (\ref{EE}).

 \subsection{Ultra-Dense Networks}
UDNs are obtained by extensively deploying diverse types of BSs in hot-spot areas and can be regarded as a massive version of heterogenous networks \cite{samarakoon2016ultra}. A general guideline for realizing green UDNS is that reducing the cell size is undoubtable the way towards high EE, but the positive effect of increasing the BS density saturates when the circuit power caused by larger amount of hardware infrastructure dominates over the transmission power.
Energy-efficient UDNs involve the following aspects.

 \textbf{1) User-centric design:} One of the green networking design principles of UDNs in 5G is the \emph{user-centric} concept where signaling and data as well as uplink and downlink transmissions are decoupled \cite{chih2014toward}.
 A well-known architecture for the decoupling of signaling and data  is the \emph{hyper-cellular structure} where macrocells provide an umbrella coverage and microcells utilizing e.g. mmWave BSs, aim at fulfilling the high capacity demands. Thereby, the significant signaling overheads caused by mobile device roaming in small size cells are reduced, which decreases the associated energy consumption at both transceivers substantially.  Furthermore, the macrocell BSs that are only capable of providing reliable coverage,  can be replaced by more energy-efficient types of BSs rather than the conventional energy-consuming ones.
 In addition, the separation of signaling and data also eases the integration of other radio access technologies (RATs), such as WiFi and future mmWave RATs, which may help to realize  further  potential EE gains. Meanwhile, decoupling downlink and uplink enables more flexible user association schemes, which also leads to substantial energy savings for both BSs and UTs. For example, for two neighbor BSs where BS 1 and BS 2  are heavily loaded and  with limited available spectrum left in the downlink and uplink, respectively, an UT can dynamically connect to BS 1 in the uplink and to BS 2 in the downlink. As such, both uplink and downlink transmit power consumption can be reduced, which is significant for 5G applications with ultra-high data rate requirements.



 \textbf{2) BS on/off switching:} Separating signaling and data  also enables efficient discontinuous transmission/reception,  i.e., BS sleeping, to save energy via exploiting the dynamics of the wireless traffic load across time and geographical locations \cite{zhang2015many}.
  It has been shown that today 80\% of BSs are quite lightly loaded for 80\% of the time but still consume almost their peak energy to account for elements such as  cooling and power amplifying circuits.  
Therefore, BS sleeping is deemed to be an effective mechanism for substantial energy savings, especially for UDNs with highly densified BSs \cite{cai2016green}.
Specifically, the data BSs (DBSs)  can be densely deployed in order to satisfy the capacity requirement during peak traffic times, while a portion of the DBSs can be switched off or put to the sleep mode if the traffic load is low in order to save energy.

\textbf{3) Interference management:} In general, cellular networks are exposed to two major sources of interference, namely, intra-cell interference and intercell interference. The former is not a significant issue in today's cellular networks due to the use of orthogonal frequency-division multiple access (OFDMA) technology and BS controlled scheduling. 
 However, the latter one is a critical concern for UDNs due to the high frequency reuse factor in multi-tier and heterogeneous networks.
 For instance, because of interference, the increases of the transmit powers of two neighboring BSs will cancel each other out without improving the system throughput, which leads to a low system EE. In addition, femtocell BSs may create ``dead zones'' around them in the downlink, especially for cell-edge users associated with other BSs.
 Therefore, efficient  interference management schemes such as power control, resource partitioning and  scheduling, cooperative transmission, and interference alignment are needed for the successful deployment of energy-efficient UDNs.  Although completely eliminating interference is overly optimistic for practical systems, it is expected that that removing the two or three strongest interferers still brings an order-wise network EE improvement.
\subsection{D2D Communications}
D2D communications \cite{feng2013device} enable densified local reuse of spectrum and can be regarded as a special case of ultra-dense networks with the smallest cell consisting of two devices as the transmitter and the receiver. In light of this, techniques that are used in UDNs may be applied to D2D scenarios.Energy-efficient D2D communications involve the following aspects.


\textbf{1) Mode selection and power control:} In D2D communication, there are two modes of communication: the cellular mode and the D2D mode.
 In the cellular mode, UTs are treated as regular cellular devices and communicate via BSs. In contrast, in the D2D mode, UTs communicate directly with each other without going through BSs by reusing the spectrum that has already been assigned to a cellular user (underlay communication) or has not been assigned to a cellular user (overlay communication). Underlay D2D communication generates co-channel interference and may switch to overlay D2D communication when the generated co-channel interference is strong.
It has been shown that underlay D2D communication is preferable for EE-oriented designs while overlay D2D communication tends to be more efficient for SE-oriented designs. This is mainly due to the interference mitigation characteristics of EE-oriented designs via limiting the transmit power.
%
%
%

%
%

\textbf{2) Active user cooperation:} With the unprecedented growth of the number of mobile devices and the data traffic,  another benefit of D2D communication is the possibility of active user cooperation, which facilitates energy savings in 5G networks, especially with regard to extending the battery lifetime of handheld devices. In particular, D2D devices can
\begin{itemize}
  \item  act as mobile relays for assisting cellular transmissions or other pairs of D2D transmissions.
  For example,   for the uplink transmission of cell-edge users, the channel conditions are generally severely degraded and direct uplink transmissions to the BS incurs exceedingly high energy consumption, which would heavily affect the user experience and satisfaction.
With proper devices as relays, significant energy consumption can be saved and it can also be further extended into multi-hop, two-way, and multiple relay scenarios.
  \item act as cluster heads for multicast transmission, e.g. for synchronous video streaming.
  It is known that if the data rate of multicast transmission is limited by the worst channel among targeted members. However,
 With D2D functionality, any member, e.g. the member with the best channel, can be selected as the cluster head and further multicasts its received data to other members, which achieves a multiuser spatial  diversity.
  Alternatively, the content may be divided into multiple chunks and each member that receives a subset of them can share its data with others.
 \item  act as local caching devices for content exchange, e.g. for asynchronous video streaming.
 Wireless caching is appealing since the data storage at the devices is often under-utilized and can be leveraged for wireless caching.
 It provides a way to exploit the inherent content reuse characteristic of 5G networks while coping with asynchronism of the demands. Devices storage is probably the cheapest and most rapidly growing network resource that, so far, has been left almost untapped for incumbent wireless networks.
 Taking into account content popularity,  deciding what content to cache plays a important role in alleviating backhaul burden as well as reducing power consumption.
\end{itemize}
Although  active user cooperation facilitates the efficient use of spectrum, it is still of practical interest to study how self-interest devices can be motivated to relay, share, and cache data for other devices at the cost of sacrificing their own limited energy. Some rewarding and pricing schemes from economic theory may be resorted to design efficient D2D cooperative protocols and mechanisms \cite{wu2016energy}.


\section{Green 5G Technologies: Enhancing spatial degrees of freedom via M-MIMO  }
M-MIMO is expected to provide at least an order of magnitude improvement in multiplexing gain and array gain via deploying a large number of antennas at the BSs while serving a large number of users with the same time-frequency resource, i.e.,  by increasing $N$ in (\ref{SE}) and (\ref{EE}). Energy-efficient M-MIMO involves  the following aspects.

\textbf{1) How many antennas are needed for green M-MIMO systems?}
For an M-MIMO system with $M$ antennas equipped at the BS and $K$ single-antenna users, it has been shown in \cite{Hien2013} that each user can scale down its uplink transmit power proportional to $M$ and $\sqrt{M}$ with perfect and imperfect channel state information (CSI), while achieving the same performance as a corresponding single-input single-output (SISO) system. However, only reducing the transmit power consumption at the users is not sufficient for improving system EE, since the overall power consumption also includes the circuit power consumption which increases linearly with the number of hardware components. 
 Basically, a general guideline towards determining the number of antennas needed for achieving green M-MIMO is: \emph{when the transmit power largely dominates the overall power consumption, deploying more antennas yields a  higher EE; while when the circuit power largely dominates the overall power consumption, further deploying more antennas is no longer energy efficient.} This is due to the  fundamental interplay between the achievable throughput gain and the additional circuit power consumed as a result of deploying more antennas.
 In light of this, a realistic power consumption model is established in \cite{bjornson2015optimal},
where it was shown that deploying $100$-$200$ antennas to serve a relatively large number of terminals is the EE-optimal solution for today's circuit technology.

\textbf{2) Signal processing and green hardware implementation:}
Centralized digital signal processing enables large savings in both hardware and power consumption. The favourable propagation conditions arising from deploying massive antennas significantly simplifies the precoding and detection algorithms at the BS. Linear signal processing techniques, such as maximum-ratio combining (MRC) and maximum-ratio transmission (MRT), can achieve a near-optimal system throughput performance.  Compared with existing sophisticated signal processing methods such as dirty-paper-coding (DPC) and  successive interference cancellation, the simplifications facilitated by M-MIMO reduce the dissipated energy required for computations.
Besides, a large number of antennas also permits green hardware implementation.
It has been shown that M-MIMO systems require a significantly lower RF transmit power, e.g. in the milliwatt range, which will result in substantial power savings in power amplifier operations.  
However, the mobile hardware at UTs is still the system performance bottleneck due to more stringent requirements on power consumption and limited physical size, which needs further investigation.



\textbf{3) Pilot design:} For M-MIMO systems,  massive timely and accurate CSI acquisition will also lead to the significant pilot power consumption since the accuracy of the channel estimation will directly affect the achievable throughput. It is known that the required pilot resources for CSI acquisition are proportional to the number of the transmit antennas, which makes  frequency-division duplexing (FDD)  unaffordable for practical implementation of M-MIMO systems and novel schemes, such as pilot beamforming and semi-orthogonal pilot design, have to be exploited.  Current research efforts mainly focus on time-division duplexing (TDD) systems, where users first send uplink pilots and then the BSs estimate the required CSI via pilots for downlink data transmission.
Furthermore, for multi-cell scenarios,  pilot contamination caused by reusing the same pilot resources in adjacent cells also affects the EE and SE significantly. Therefore, how to balance the the use of the  time-frequency resource for pilot training and data transmission in both uplink and downlink as well as designing low complexity pilot mitigation schemes are important challenges for achieving high EE in M-MIMO systems.

\section{Enabler of Sustainable 5G Communications: Energy Harvesting}
Beside the above techniques for improving the EE, the energy harvesting technologies drawing energy from renewable resources or RF signals are important enablers of sustainable green 5G wireless networks. In Table \ref{table2},  energy harvesting from renewable resources and RF energy harvesting are compared from different perspectives.

\subsection{Energy Harvesting from Renewable Resources}
Renewable resources energy harvesting transforms the natural energy sources such as solar, wind, etc. into electrical energy and has been
recognized as a promising green technology  to alleviate the energy crisis and to reduce the energy costs.
The following research directions for communication system design based on renewable energy have emerged.
\begin{table*}[!t]
\caption{Comparison of energy harvesting technologies.}\label{table2}
\centering
\newcommand{\tabincell}[2]{\begin{tabular}{@{}#1@{}}#2\end{tabular}}
\small
\begin{tabular}{|p{4.8cm}|p{2.23cm}|p{1.3cm}|p{3cm}|p{3cm}|}
  \hline
Technology & Energy Source Controllability& Efficiency & Ultimate Goal & 5G Applications\\ \hline
Natural Resource Energy Harvesting &  Non-controllable &-  &Reduce oil dependency& Hybrid BSs\\ \hline
   \tabincell{l}{RF Energy Harvesting} & Controllable  & Low &  Sustainable devices& IoT\\ \hline
\end{tabular}
\end{table*}

\textbf{1) Distinctive energy-data causality and battery capacity constraints.}
A fundamental feature that distinguishes energy harvesting communication systems from non-energy harvesting communication systems is the existence of an energy causality constraint \cite{gunduz2014designing}. This constraint imposes a restriction on the energy utilization, i.e.,  energy cannot be utilized before it is harvested, which is in essence attributed to intermittent availability of renewable resources.  In order to  fully embrace the potential benefits of energy harvesting, carefully incorporating this constraint into the design  5G networks and gaining deep understanding of its impacts  on 5G networks are vital.
Two other relevant constraints are the data causality constraint and the battery capacity constraint, which depend on the data traffic variations and energy storage limitations. Specifically, the data causality constraint accounts for the fact that  a data packet cannot be delivered before it has arrived at the transmitter.
The battery capacity constraint models the limited capacity of the battery for storing harvested energy.
Thus, the conventional research studies assuming infinitely  backlogged data buffers and  unlimited energy storages  can be regarded as special cases of the above constraints.  However, in upcoming 5G networks, these assumptions  are no longer valid due to  the extremely diverging traffic and  device types. Thus, new QoS concerns on energy harvesting communication  naturally arise in 5G systems.




\textbf{2) Offline versus online approaches:} Depending on whether the energy arrivals, CSI, data traffic, etc., are predictable or not, resource allocation algorithms for communication systems with energy
harvesting can be categorized into \emph{offline} and \emph{online} policies.  Specifically, based on non-causal knowledge information,
optimal offline solutions can be developed by exploiting tools from optimization theory. Although this approach is  impractical  due to  highly dynamic network variations in 5G communication networks, offline solutions provide performance upper bounds for online algorithms.
 In contrast, based on causal information,  online schemes can be developed and  further categorized into two branches. The first branch assumes that statistical knowledge information is available at the transmitter and the optimal solutions thereby can be numerically obtained by exploiting dynamic programming theory. Yet, such an approach lacks of analytical insight and cannot be exploited for handling large scale system due to the ``curse of dimensionality". Alternatively, some suboptimal online algorithms are desirable in practice, which can be designed based on the insights observed from the optimal offline results. However, in some of the 5G application scenarios, statistical characteristics may change over time, or such information may not be available before deployment. This leads to a more aggressive branch that does not require any priori information at the transmitter. The optimal solutions in this branch rely on the learning theory, where the transmitter learns the optimal transmission policy over time by iteratively performing actions and observing their immediate rewards.


\textbf{3) Hybrid energy supply solutions and energy cooperation.}
Due to the uncertain and dynamic changing environmental conditions, the harvested energy is generally intermittent by nature.
This poses new challenges for powering 5G systems that require stable and satisfactory QoS.
A pragmatic approach is to design a hybrid energy harvesting system, which uses different energy sources, e.g. solar, wind, diesel generators, or even the power grid, in a complementary manner to provide uninterrupted service.
%
%
%
As an example,  for 5G communication scenarios such as UDNs with a massive number of BSs, the integration of energy harvesting not only provides location diversity to compensate for the additional circuit power consumption caused by deploying more BSs but also reduces the power grid power consumption significantly.  In addition, densified BSs employing energy harvesting also facilitate possible energy cooperation between BSs, which has huge potential for realizing green 5G networks. For example, the BSs located in ``good'' spots will harvest more energy and can correspondingly serve more users or transfer their superfluous energy through power lines to BSs that harvest less energy to realize energy cooperation.

\subsection{RF Energy Harvesting}
RF energy harvesting, also known as wireless power transfer (WPT), allows receivers to harvest energy from received RF signals \cite{lu2015wireless,bi2015wireless}. Compared to opportunistic energy harvesting from renewable resources,  WPT can be fully controlled via wireless communication technologies, which has been thereby regarded as a promising solution for powering the massive amount number of small UTs expected in 5G applications such as IoT.
 The following research directions for RF energy harvesting have been identified.

\textbf{1) Typical system architectures and extensions:}
In the literature, two canonical lines of research can be identified for implementing WPT. 
The first line focuses on characterizing the fundamental tradeoff between the achievable rate and the harvested energy by exploring the dual use of the same signal.
This concept is known as simultaneous  wireless information and power transfer (SWIPT). Specifically, wireless devices split the received signal sent by a BS into two parts, one for information decoding and the other for energy harvesting. Thus, a fundamental issue of this line of work is to optimize the power splitting ratio at the receiver side to customize the achievable throughput and the harvested energy. The other line of research separates energy harvesting and information transmission in time, i.e., harvest and then transmit, leading to wireless powered communication networks (WPCNs). Specifically, wireless devices first harvest energy from signals sent by a power station and then perform wireless information transmission (WIT) to the target BS/receiver using harvested energy. Thus, the WPT and WIT durations have to be carefully designed,  since more time for WPT leads to more harvested energy for using in WIT but leaves less time for WIT.
\emph{These two canonical architectures as well as their extensions serve as a foundation for wireless-powered UDNs, M-MIMO,  mmWave networks, etc., which utilize widely RF energy flow among different entities of 5G networks.}

\textbf{2) Improving the efficiency of WPT:}
 Despite the advantages of WPT, the system performance is basically constrained by the low energy conversion efficiency and the severe path loss during WPT \cite{krikidis2014simultaneous}.
Generally, the 5G  techniques that are able to improve the performance of a wireless communication link can also be exploited to improve the efficiency of WPT.
 For example, narrow beams can be realized by employing multiple antennas at the transmitter side with optimized beamforming vectors, which is known as \emph{energy beamforming} \cite{bi2015wireless} and this fits well the applications of M-MIMO and mmWave. In addition, short range communication based 5G techniques such as  D2D communications and UDNs are capable of improving the efficiency of WPT by reducing the energy transfer distance.  Besides improving the amount of harvested energy, the circuit power consumed by information decoding is also an important design issue since it reduces the net harvested energy that can be stored in the battery for future use.
 In particular, the active mixers used in conventional information receivers for RF to baseband conversion are substantially power-consuming. This motivates additional efforts on designing novel receiver architectures that consume less power by avoiding the use of active devices.

 \textbf{3) Rethinking interference in wireless networks with WPT:} Another advantage of WPT is that not only dedicated signals but also co-channel interference signals can be exploited for energy harvesting.  This fundamentally shifts our traditional viewpoint of ``harmful'' co-channel interference, which now becomes a potential source of energy. In this regard, UDNs and D2D underlaying communications where the spectrum is heavily reused across the whole network, provide plentiful opportunities for WPT  to extract the potential gain from co-channel interference.
  In practice, WPT enabled devices can harvest energy when co-channel interference is strong and decode information when co-channel interference is relatively weak.  Besides, one may inject artificial interference deliberately into communication systems, which may be beneficial for the overall system performance, especially when receivers are hungry for energy to support its normal operations. As such, this paradigm shift provides new degrees of freedom to optimize the network interference levels for achieving the desired tradeoff between information transmission and energy transfer.

\section{Conclusions}
\begin{figure*}[!t]
\setlength{\abovecaptionskip}{0pt}
\setlength{\belowcaptionskip}{0pt}
\centering
\includegraphics[width=1\textwidth]{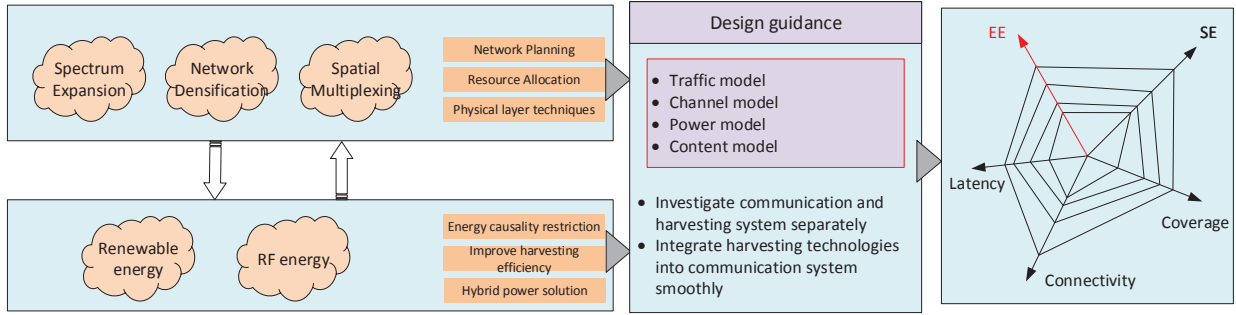}
\setlength{\belowcaptionskip}{0pt}
\caption{An overview of components of sustainable green 5G networks. }\label{summary}
\end{figure*}

In this article, we have surveyed the advanced technologies which are expected to enable sustainable green 5G networks. A holistic design overview can be found in Fig. \ref{summary}. Energy harvesting underpins the green expectation towards 5G networks while promising spectrum-efficient 5G technologies can be tailored to realize energy-efficient wireless networks. Facing the highly diversified communication scenarios of future, user traffic, channel,  power consumption, and  even content popularity models need to be jointly taken into account for improving the system EE. Thereby, it is evident that the diverse applications and heterogeneous user requirements of sustainable green 5G networks cannot be satisfied with any particular radio access technology. Instead, an ecosystem of  interoperable technologies is needed such that the respective technological advantages of the different components,  can be exploited together pushing towards the ultimate performance limits. This, however, also poses new challenges for the system designers.

\ifCLASSOPTIONcaptionsoff
  \newpage
\fi


\bibliographystyle{IEEEtran}
\bibliography{IEEEabrv,mybib}

\begin{thebibliography}{10}
\providecommand{\url}[1]{#1}
\csname url@samestyle\endcsname
\providecommand{\newblock}{\relax}
\providecommand{\bibinfo}[2]{#2}
\providecommand{\BIBentrySTDinterwordspacing}{\spaceskip=0pt\relax}
\providecommand{\BIBentryALTinterwordstretchfactor}{4}
\providecommand{\BIBentryALTinterwordspacing}{\spaceskip=\fontdimen2\font plus
\BIBentryALTinterwordstretchfactor\fontdimen3\font minus
  \fontdimen4\font\relax}
\providecommand{\BIBforeignlanguage}[2]{{%
\expandafter\ifx\csname l@#1\endcsname\relax
\typeout{** WARNING: IEEEtran.bst: No hyphenation pattern has been}%
\typeout{** loaded for the language `#1'. Using the pattern for}%
\typeout{** the default language instead.}%
\else
\language=\csname l@#1\endcsname
\fi
#2}}
\providecommand{\BIBdecl}{\relax}
\BIBdecl

\bibitem{andrews2014will}
J.~G. Andrews, S.~Buzzi, W.~Choi, S.~V. Hanly, A.~Lozano, A.~C. Soong, and
  J.~C. Zhang, ``What will {5G} be?'' \emph{{IEEE} J. Sel. Areas Commun.},
  vol.~32, no.~6, pp. 1065--1082, Jun. 2014.

\bibitem{chen2011fundamental}
Y.~Chen, S.~Zhang, S.~Xu, and G.~Y. Li, ``Fundamental trade-offs on green
  wireless networks,'' \emph{{IEEE} Commun. Mag.}, vol.~49, no.~6, pp. 30--37,
  Jun. 2011.

\bibitem{gunduz2014designing}
D.~Gunduz, K.~Stamatiou, N.~Michelusi, and M.~Zorzi, ``Designing intelligent
  energy harvesting communication systems,'' \emph{{IEEE} Commun. Mag.},
  vol.~52, no.~1, pp. 210--216, Jan. 2014.

\bibitem{han2015large}
S.~Han, I.~Chih-Lin, Z.~Xu, and C.~Rowell, ``Large-scale antenna systems with
  hybrid analog and digital beamforming for millimeter wave {5G},''
  \emph{{IEEE} Commun. Mag.}, vol.~53, no.~1, pp. 186--194, Jan. 2015.

\bibitem{zhang2015lte}
R.~Zhang, M.~Wang, L.~X. Cai, Z.~Zheng, X.~Shen, and L.-L. Xie,
  ``{LTE}-unlicensed: the future of spectrum aggregation for cellular
  networks,'' \emph{{IEEE} Wireless Commun. Mag.}, vol.~22, no.~3, pp.
  150--159, Jun. 2015.

\bibitem{samarakoon2016ultra}
S.~Samarakoon, M.~Bennis, W.~Saad, M.~Debbah, and M.~Latva-aho, ``Ultra dense
  small cell networks: Turning density into energy efficiency,'' \emph{{IEEE}
  J. Sel. Areas Commun.}, vol.~34, no.~5, pp. 1267--1280, Apr. 2016.

\bibitem{chih2014toward}
I.~Chih-Lin, C.~Rowell, S.~Han, Z.~Xu, G.~Li, and Z.~Pan, ``Toward green and
  soft: a {5G} perspective,'' \emph{{IEEE} Commun. Mag.}, vol.~52, no.~2, pp.
  66--73, Feb. 2014.

\bibitem{zhang2015many}
S.~Zhang, J.~Gong, S.~Zhou, and Z.~Niu, ``How many small cells can be turned
  off via vertical offloading under a separation architecture?'' \emph{{IEEE}
  Trans. Wireless Commun.}, vol.~14, no.~10, pp. 5440--5453, Oct. 2015.

\bibitem{cai2016green}
S.~Cai, Y.~Che, L.~Duan, J.~Wang, S.~Zhou, and R.~Zhang, ``Green {5G}
  heterogeneous networks through dynamic small-cell operation,'' \emph{{IEEE}
  J. Sel. Areas Commun.}, vol.~34, no.~5, pp. 1103--1115, May 2016.

\bibitem{feng2013device}
D.~Feng, L.~Lu, Y.~Yuan-Wu, G.~Y. Li, G.~Feng, and S.~Li, ``Device-to-device
  communications underlaying cellular networks,'' \emph{{IEEE} Trans. Commun.},
  vol.~61, no.~8, pp. 3541--3551, Aug. 2013.

\bibitem{wu2016energy}
Q.~Wu, G.~Y. Li, W.~Chen, and D.~W.~K. Ng, ``Energy-efficient small cell with
  spectrum-power trading,'' \emph{arXiv preprint arXiv:1608.03178}, 2016.

\bibitem{Hien2013}
H.~Q. Ngo, E.~Larsson, and T.~Marzetta, ``Energy and spectral efficiency of
  very large multiuser {MIMO} systems,'' \emph{{IEEE} Trans. Commun.}, vol.~61,
  no.~4, pp. 1436--1449, Apr. 2013.

\bibitem{bjornson2015optimal}
E.~Bj{\"o}rnson, L.~Sanguinetti, J.~Hoydis, and M.~Debbah, ``Optimal design of
  energy-efficient multi-user {MIMO} systems: Is massive {MIMO} the answer?''
  \emph{{IEEE} Trans. Wireless Commun.}, vol.~14, no.~6, pp. 3059--3075, Jun.
  2015.

\bibitem{lu2015wireless}
X.~Lu, P.~Wang, D.~Niyato, D.~I. Kim, and Z.~Han, ``Wireless networks with {RF}
  energy harvesting: A contemporary survey,'' \emph{{IEEE} Commun. Surveys
  Tuts.}, vol.~17, no.~2, pp. 757--789, Second Quart 2015.

\bibitem{bi2015wireless}
S.~Bi, C.~K. Ho, and R.~Zhang, ``Wireless powered communication: opportunities
  and challenges,'' \emph{{IEEE} Commun. Mag.}, vol.~53, no.~4, pp. 117--125,
  Apr. 2015.

\bibitem{krikidis2014simultaneous}
I.~Krikidis, S.~Timotheou, S.~Nikolaou, G.~Zheng, D.~W.~K. Ng, and R.~Schober,
  ``Simultaneous wireless information and power transfer in modern
  communication systems,'' \emph{IEEE Commun. Mag.}, vol.~52, no.~11, pp.
  104--110, Nov. 2014.

\end{thebibliography}

%
%
%

\end{document}